\documentclass{amsart}
\usepackage[english]{babel}
\usepackage{graphicx} 
\usepackage{amsmath, amsthm, latexsym, amssymb, amsfonts, epsf, xcolor}

\usepackage[hidelinks]{hyperref}
\usepackage[biblabel]{cite}

\newtheorem{theorem}{Theorem}[section]
\newtheorem{lemma}[theorem]{Lemma}
\newtheorem{proposition}[theorem]{Proposition}

\theoremstyle{definition}
\newtheorem{definition}[theorem]{Definition} 
\newtheorem{remark}[theorem]{Remark}

\usepackage{mathtools} 
\DeclarePairedDelimiter\abs{\lvert}{\rvert}%
\DeclarePairedDelimiter\norm{\lVert}{\rVert}%

\makeatletter
\let\oldabs\abs
\def\abs{\@ifstar{\oldabs}{\oldabs*}}
\let\oldnorm\norm
\def\norm{\@ifstar{\oldnorm}{\oldnorm*}}
\makeatother

\DeclareMathOperator{\ev}{ev}
\DeclareMathOperator{\rk}{rk}
\DeclareMathOperator{\wt}{wt}
\DeclareMathOperator{\GRS}{GRS}
\DeclareMathOperator{\Hull}{Hull^{H}}

\newcommand{\F}{{\mathbb{F}}}
\newcommand{\fq}{\mathbb{F}_q}
\newcommand{\cL}{\mathcal{L}}
\newcommand{\cT}{\mathcal{T}}
\newcommand{\cF}{\mathcal{F}}
\newcommand{\cB}{\mathcal{B}}

\newcommand{\cP}{\mathcal{P}}
\newcommand{\C}{C_{\lambda,\tau,\rho,\sigma}(k)}

\newcommand{\specialcell}[2][c]{%
\begin{tabular}[#1]{@{}c@{}}#2\end{tabular}}

\begin{document}
\title[Hermitian hull of some GRS codes and new EAQMDS codes]{Hermitian hull of some GRS codes and new entanglement-assisted quantum MDS codes}

\author[O. Campion]{Oisin Campion}
\address[Oisin Campion]{School of Mathematics and Statistics, University College Dublin, Ireland}
\email{oisin.campion@ucdconnect.ie}

\author[R. San-José]{Rodrigo San-José}
\address[Rodrigo San-José]{IMUVA-Mathematics Research Institute, Universidad de Valladolid, 47011 Valladolid (Spain).}
\email{rsanjose@vt.edu}
\email{rodrigo.san-jose@uva.es}

\thanks{This publication has emanated from research conducted with the financial support of Science Foundation Ireland under Grant number 21/RP-2TF/10019 for the first author. The second author has been partially supported by Grant PID2022-138906NB-C21 funded by MICIU/AEI/ 10.13039/501100011033 and by ERDF/EU, and by Grant FPU20/01311 funded by the Spanish Ministry of Universities.}

\begin{abstract}
We study the Hermitian hull of a particular family of generalized Reed-Solomon codes. The problem of computing the dimension of the hull is translated to a counting problem in a lattice. By solving this problem, we provide explicit formulas for the dimension of the hull, which determines the minimum number required of maximally entangled pairs for the associated entanglement-assisted quantum error-correcting codes. This flexible construction allows to obtain a wide range of entanglement-assisted quantum MDS codes, as well as new parameters. 
\end{abstract}

\maketitle

\section{Introduction}
Let $\fq$ be a finite field with cardinality $q$, where $q$ is a prime power, and let $n$ be a positive integer. An $[n,k,d]_q$ linear code $C\subset \fq^n$ is a $k$-dimensional linear subspace of $\fq^n$, with minimum distance $d$. The hull of a linear code, defined as the intersection of $C$ with its dual (for example, with respect to the Euclidean or Hermitian inner product), has received a lot of attention recently. In particular, there has been work related to the study of the possible dimensions of the hulls using equivalent codes \cite{relativehull,haochenhull,carletHullVariation,grasslvariableentanglement}, and also there has been some work related to the computation of the hull for certain families of codes \cite{luoHullSimplex,luoHermitianHullsMDS,kaplanHullsPRM,sanjoseHullsPRM,sanjoseHullvariationPRM}. The interest in studying the hull comes from the applications in several areas, such as determining the automorphism group of linear codes \cite{leonComputingAutomorphismGroups}, code equivalence \cite{sendriePermutationEquivalentCodes}, or entanglement-assisted quantum error-correcting codes (EAQECCs) \cite{brunentanglement,galindoentanglement}. 

The Singleton bound for linear codes states that $d\leq n-k+1$. A code is called an MDS code if $d=n-k+1$, that is, if we have equality in the Singleton bound. Generalized Reed-Solomon (GRS) codes are MDS codes that are obtained by evaluating one-variable polynomials at points of a finite field $\fq$. GRS codes are among the most well-known families of linear codes, and the study of their hulls is a topic of current interest \cite{chenLargeHermitianHull,wuGaloisHullGRS,fangGaloisHullsGRS,gaoHullsGRSviaGoppa,chenHullsRSviaAG,huangDimHullGRS}. In this paper, we are interested in the Hermitian hull of a particular class of GRS codes and its application to EAQECCs. 

There has been significant interest in quantum computing in recent years due to the existence of quantum algorithms outperforming classical ones for certain tasks, e.g., see \cite{shorPrimeFactorization}. However, quantum error-correction is needed to guard against noise and decoherence. The independent works \cite{calderbankgoodquantum,cssoriginal2} showed how to use classical codes to construct quantum error-correcting codes (QECCs), which is the so-called CSS construction. An extension of these codes is given by EAQECCs, making use of pre-existing entanglement between transmitter and receiver to increase the transmission rate \cite{brunentanglement}. EAQECCs can also be constructed with classical codes \cite{galindoentanglement}, and determining their performance requires the computation of one extra parameter $c$, which is the minimal number of maximally entangled pairs required. This value is determined by the dimension of the hull of the classical code used, motivating the study of the hulls of classical codes. EAQECCs also satisfy a Singleton-type bound, e.g., see \cite{grasslSingletonEAQECC}, and the EAQECCs that achieve equality in this bound are called entanglement-assisted quantum MDS codes (EAQMDS, or QMDS if they do not require entanglement assistance). In the classical setting, for every set of parameters satisfying the Singleton bound, we know how to construct an MDS code with those parameters, provided that $n\leq q+1$. In the quantum setting, it is conjectured that the maximum length of an EAQMDS is $q^2+1$ (besides some exceptions), but, unlike in the classical setting, we do not have constructions for every set of parameters allowed by the quantum Singleton bound, even if we assume $n\leq q^2+1$, e.g., see \cite{grasslQMDSSmallFields}. It is thus desirable to find EAQMDS codes whose parameters could not be achieved with previous constructions.

In this paper, we consider a flexible family of GRS codes, which was introduced in \cite{campionQMDS}. This family provides QMDS codes with new parameters. From QMDS codes, one can consider propagation rules \cite{grasslvariableentanglement} to derive EAQMDS codes. For example, this is done in \cite{chenEAQMDS} for some QMDS codes. However, this approach is quite limited, for example, in terms of the minimum distance that can be achieved. Therefore, in this paper we consider codes with higher minimum distance, which are no longer self-orthogonal with respect to the Hermitian inner product, and we compute their Hermitian hull. This problem translates to a counting problem in a lattice, which we solve explicitly to completely determine the parameters of the corresponding quantum codes. 

The content of the paper is organized as follows. In Section \ref{s:prelims}, we describe a flexible family of generalized Reed-Solomon codes and provide some basic facts about their Hermitian hulls. In Section \ref{s:orthog} we derive orthogonality conditions for our codewords and translate the problem of computing the dimension of the Hermitian hull into counting the number of points on certain lattices. These types of lattices are studied in their generic form in Section \ref{s:countinglattices}, which allows us to define and study the relevant lattices for our codes in Sections \ref{s:laticeT} and \ref{s:latticeP}. In Section \ref{s:computec} we give explicit formulas that can be used to calculate the dimension of the Hermitian hull of our codes, and in Section \ref{s:comparison} we compare the resulting EAQECCs with the best know codes in the literature. 

\section{Preliminaries}\label{s:prelims}
We start by defining GRS codes. Fix $A=\{a_1,\dots,a_n\}\subset \F_{q^2}^n$ and $v\in(\F_{q^2}^*)^n$. For $1\leq k \leq n$, we consider $\F_{q^2}[X]_{<k}$, that is, the univariate polynomials over $\F_{q^2}$ of degree less than $k$. We define the evaluation map
$$
\ev_{v,A}:\F_{q^2}[X]_{<k} \rightarrow \F_{q^2}^n,\; f\mapsto (v_1f(a_1),\dots,v_nf(a_n)). 
$$
\begin{definition}
The GRS code $\GRS_{n,k}(v,A)$ is defined as 
$$
\GRS_{n,k}(v,A):=\ev_{v,A}(\F_{q^2}[X]_{<k}).
$$
\end{definition}
The parameters of $\GRS_{n,k}(v,A)$ are $[n,k,n-k+1]_q$, which means that these codes are MDS. The dual of a GRS code is also a GRS code, with parameters $[n,n-k,k+1]_q$. 

Now we define the particular family of GRS codes we consider, which was already introduced in \cite{campionQMDS}. For this, we will first define a particular set $A$, and then a particular vector $v$. Assume now that $q\geq 4$. Let $\lambda>1$ be a divisor of $q-1$, and 
let $\tau>1$ and  $\rho>1$ be divisors of $q+1$. We  assume that $\gcd (\lambda,\tau)=1$.
We let  $\kappa_1=\gcd (\lambda,\rho), \kappa_2= \gcd(\tau,\rho)$, and $ \kappa=\kappa_1\kappa_2$. Let $n=\lambda \tau \sigma$. We assume that $ \frac{\rho}{\kappa} \ge 2$, and we let $\sigma$ be any integer with $ \frac{\rho}{\kappa} \ge  \sigma \ge 2$. We denote by $\zeta_t$ a primitive $t$-th root of unity.

We consider the set
\[
    A := \{\zeta_\lambda^{i}\zeta_{\tau}^{j}\zeta_{\rho}^{\ell} : 
    0\leq i < \lambda, \ 0\leq j < \tau,\
    0\leq \ell < \sigma\} \subset \F_{q^2}.
\]

By \cite[Lem. 3.1]{campionQMDS}, the elements of $A$ are distinct, and we can uniquely associate triples $(i,j,\ell)$ with elements of $A$, defining $A(i,j,\ell):=\zeta_\lambda^{i}\zeta_{\tau}^{j}\zeta_{\rho}^{\ell}$. Now choose $s_0,\ldots s_{\sigma-1}\in \F_q^*$ in the following way: 
\begin{itemize}
    \item If $\sigma=2$, then set $s_0=1, s_1=-1$. 
    \item Otherwise, set $s_0,\ldots,s_{\sigma-3} =1$, select $s_{\sigma-2} \in \F_q $ different from $ \{0,-(s_0+\ldots +s_{\sigma-3}), -(s_0+\ldots +s_{\sigma-3})/2\}$ (if the latter exists), and set $s_{\sigma-1} = -(s_0+\ldots s_{\sigma-2})$. This requires the assumption $q\ge 4$. 
\end{itemize}

This ensures that $\sum_{\ell=0}^{\sigma-1}s_\ell=0$ and that $s_{\sigma-2} \neq s_{\sigma-1}$ for $\sigma>2$. In \cite{campionQMDS}, another parameter $L$ is considered, which is then appropriately chosen to maximimize the range of parameters in which the codes we will consider are self-orthogonal. For this work, we will only consider the optimal values of $L$ obtained in \cite{campionQMDS}, which are listed in Section \ref{s:laticeT} in Table \ref{tab:valueL}.

We can define now the vector $v\in (\F_{q^2}^*)^n$. We denote by $v(i,j,\ell)$ the coordinate of $v$ associated to $A(i,j,\ell)$, and we consider $v(i,j,\ell)\in \F_{q^2}$ such that
$$
v(i,j,\ell)^{q+1}:=\zeta_{\lambda}^{-i L} s_{\ell}.
$$
Since $\zeta_{\lambda}^{-i L} s_{\ell}\in \fq$, it is always possible to find such $v(i,j,\ell)$ (recall the properties of the norm map from $\F_{q^2}$ to $\fq$). With these definitions, we denote
$$
\C:=\GRS_{n,k}(v,A).
$$
Its parameters are $[\lambda\tau\sigma,k,\lambda\tau\sigma-k+1]_{q^2}$, and the parameters of its dual are  $[\lambda\tau\sigma,\lambda\tau\sigma-k,k+1]_{q^2}$.

We now introduce the construction of EAQECCs we will use. Let $u,w\in \F_{q^2}^n$. Their Hermitian inner product is
$$
u\cdot_hw:=\sum_{i=1}^nu_iw_i^q.
$$
Given $C\subset \F_{q^2}^n$, we consider its Hermitian dual
$$
C^{\perp_h}:=\{u\in \F_{q^2}^n:u\cdot_h c=0, \textnormal{ for all }c\in C\}.
$$
It is not hard to check that the parameters of $C^{\perp_h}$ and $C^\perp$ are the same, since $C^{\perp_h}=(C^\perp)^q$, where $(C^\perp)^q$ is obtained by taking the $q$-th power of the entries of the vectors in $C^\perp$. The Hermitian hull is then defined as
$$
\Hull(C):=C\cap C^{\perp_h}.
$$

The following result can be found in \cite{galindoentanglement}. 

\begin{theorem}[Hermitian construction]\label{t:hermitian}
Let $C\subset \F_{q^2}^n$ be a linear code of dimension $k$ and $C^{\perp_h}$ its Hermitian dual. Then, there is an EAQECC with parameters $[[n,K,d;c]]_q$, where
$$
c=k-\dim(\Hull(C)), \;K=n-2k+c, \; \text{ and } \;d=\wt(C^{\perp_h}\setminus \Hull(C)).
$$
\end{theorem}

The Hermitian hull of $C$ is defined as $C\cap C^{\perp_h}$. From the previous result we see that computing the dimension of the Hermitian hull of $C$ is equivalent to finding the parameter $c$. When $C\cap C^{\perp_h}=C$, the code is Hermitian self-orthogonal, and we recover the usual Hermitian construction without entanglement assistance \cite{kkks}. 

As stated in the introduction, $c$ is the minimum number required of maximally entangled qudit pairs required, and it can we rewritten as
$$
c=\dim C-\dim C\cap C^{\perp_h}=\rk G\cdot (G^q)^t,
$$
where $G^q$ is the matrix whose entries are the $q$-th power of the entries of the generator matrix $G$ of $C$. Now consider $C=\C$. Let $g_i$, $1\leq i \leq k$, be the rows of $G$. Then we can assume that $g_i=\ev_{v,A}(X^{i-1})$, and then 
\begin{equation}\label{eq:entries}
(G\cdot (G^q)^t)_{i,j}=\ev(X^{i-1})\cdot_h\ev(X^{j-1}).
\end{equation}

We have the following Singleton bound for EAQECCs from \cite[Cor. 9]{grasslSingletonEAQECC}. 

\begin{theorem}\label{t:singleton}
Consider an EAQECC with parameters $[[n,k,d;c]]_q$. Then
$$
\begin{aligned}
    &k\leq c+\max\{0,n-2d+2\},\\
    &k\leq n-d+1,\\
    &k\leq \frac{(n-d+1)(c+2d-2-n)}{3d-3-n} \; \textnormal{ if } d-1\geq \frac{n}{2}.
\end{aligned}
$$
\end{theorem}

In the next sections we will compute the Hermitian hull of the codes $\C$, and we will obtain codes that are optimal with respect to the bounds from the previous result, that is, they are EAQMDS.

\section{Orthogonality conditions and lattices}\label{s:orthog}

We start the study of $\Hull(\C)$ by determining when the evaluation of two monomials is orthogonal with respect to the Hermitian inner product. 

\begin{lemma}\label{l:sumsubroup}
Let $N\geq 0$, and $\gamma>0$ such that $\gamma \mid q^2-1$. We have the following:
$$
\sum_{i=0}^{\gamma-1}\zeta_\gamma^{iN}=
\begin{cases}
0 &\text{ if } N\not\equiv 0 \bmod \gamma,\\
\gamma  &\text{ if } N\equiv 0 \bmod \gamma.
\end{cases}
$$
\end{lemma}
\begin{proof}
The result for $N\not\equiv 0 \bmod \gamma$ follows from the formula for the sum of a geometric series, and for the case $N\equiv 0 \bmod \gamma$ we get $\sum_{i=0}^{\gamma-1}1=\gamma$.
\end{proof}

\begin{proposition}\label{p:orthogconditions}
    Let $X^{e_1},X^{e_2}$ be two monomials. Then
    $$
    ev_{v,A}\left(X^{e_1}\right)\cdot_h ev_{v,A}\left(X^{e_2}\right) =0
    $$
    if 
    any one of the following conditions holds: 
    \begin{itemize}
    \item $e_1+e_2  \not\equiv L $ (mod $\lambda$).
    \item $e_1 \not\equiv e_2$ (mod $\tau$).
    \item $e_1\equiv e_2$ (mod $\rho$).
\end{itemize}
Moreover, if $\sigma=2, 3 $ or $\rho$, then the set of conditions is both necessary and sufficient.
\end{proposition}
\begin{proof}
    The Hermitian inner product of the evaluation vectors is given by

\begin{equation*}
\begin{split}
    \ev_{v,A}\left(X^{e_1}\right)\cdot_h \ev_{v,A}\left(X^{e_2}\right) = & \sum_{i,j,\ell}\boldsymbol{v}(i,j,\ell)^{q+1} A(i,j,\ell)^{e_1 +qe_2}\\
    =&\sum_{i,j,\ell}\zeta_{\lambda}^{-i L}  s_{\ell} \ (\zeta_\lambda^{i}\zeta_{\tau}^{j}\zeta_{\rho}^{\ell})^{e_1+qe_2}\\
    =&\left(\sum_{i=0}^{\lambda-1}\zeta_{\lambda}^{i(e_1+qe_2-L)}\right)\left(\sum_{j=0}^{\tau-1}\zeta_{\tau}^{j(e_1+qe_2)}\right)\left(\sum_{\ell=0}^{\sigma-1}s_{\ell}\zeta_{\rho}^{\ell(e_1+qe_2)}\right).\\
\end{split}  
\end{equation*}

It is easy to tell exactly when the first two terms are zero using Lemma \ref{l:sumsubroup}:
\begin{itemize}
    \item The first term is zero $\iff$ $e_1+qe_2-L\not\equiv 0$ (mod $\lambda$) $\iff$ 
    $e_1+e_2  \not\equiv L $ (mod $\lambda$). 
    \item The second term is zero $\iff$ $e_1+qe_2\not\equiv 0$ (mod $\tau$) 
    $\iff$ $e_1 \not\equiv e_2$ (mod $\tau$).
\end{itemize}

We must determine under what circumstances the third term is zero. Let $\omega:= \zeta_{\rho}^{(e_1+qe_2)}$, so that the third term is $\sum_{\ell=0}^{\sigma-1}s_{\ell}\omega^\ell$. Since $\sum_{\ell=0}^{\sigma-1}s_\ell=0$, it is clear that if $\omega=1$, the sum vanishes. This occurs precisely when $e_1\equiv e_2$ (mod $\rho$). We now analyze if the sum can vanish for $w\neq 1$.

First, consider the case with $\sigma=2$. Then the third term is simply $1-\omega$, which vanishes $\iff$ $\omega =1$. 

Next, consider the case with $\sigma =3$ and suppose that $s_0+s_1\omega +s_2\omega^2=0$ with $\omega \neq 1$. Subtract the equation $s_0+s_1+s_2=0$ to get that $s_1(\omega-1)+s_2(\omega^2-1)=0$, which implies that $s_1+s_2(\omega+1)=0$, If $\omega=-1$, this implies that $s_1=0$, a contradiction. If $w \neq -1$ then  $\omega=-s_1/s_2-1 \in \F_q$. However, this would mean that the order of $\omega$ divides both $(q-1)$ and $(q+1)$, which can only happen if $\omega=1$ or $-1$, a contradiction. Thus, the only way for the sum to be zero is with $\omega=1$. 

Finally, we analyze the case $\sigma = \rho>2$. Let $t$ be the order of $\omega$ and first suppose that $t\ge 3$. This means that $t \nmid q-1$, so $\omega \notin \F_q$. 
Supposing the third term is zero, we can rewrite the sum as 
\[
\sum_{\ell=0}^{\rho-1}s_\ell\omega^\ell = \sum_{\ell=0}^{\rho-3}\omega^\ell +s_{\rho-2}\omega^{\rho-2} + s_{\rho-1}\omega^{\rho-1}. 
\]
By making the substitution $\sum_{\ell=0}^{\rho-3}\omega^\ell=-(\omega^{\rho-2}+\omega^{\rho-1})$ (see Lemma \ref{l:sumsubroup}) and dividing across by $\omega^{\rho-2}$, we get the following equation: 
\[
\omega(s_{\rho-1}-1)= 1-s_{\rho-2}
\]

If $s_{\rho-1}=1$, then we must have $s_{\rho-2}=1$. This would mean that $\sum_{\ell=0}^{\rho-1}s_\ell=\rho\cdot1= 0$, but $\rho\cdot1\neq0$  since the characteristic cannot divide $\rho$, so in fact $s_{\rho-1}\neq 1$.

From this it follows that $\omega \in \F_q$, which is a contradiction. So the sum cannot be zero for $t\ge 3$. 

Now suppose that $t=2$. This can only happen if the characteristic is different from 2, and it means that $\omega=-1$, and that $\rho$ is even. Now supposing that our third term is zero, we find that 
\[
0=\sum_{\ell=0}^{\rho-1}s_\ell(-1)^\ell = \sum_{\ell=0}^{\rho-3}(-1)^\ell +s_{\rho-2}\omega^{\rho-2} + s_{\rho-1}\omega^{\rho-1} = 0-s_{\rho-2}+s_{\rho-1}
\]
which means that $s_{\rho-1}=s_{\rho-2}$, again a contradiction. 

\end{proof}

The previous result motivates the following definition.

\begin{definition}\label{def_failurepoint}
    Let $X^{e_1},X^{e_2}$ be two monomials. If all three of the following conditions hold, then we call  the point $\left(e_1,e_2\right)$ a \textbf{failure point}: 
    \begin{equation}\label{eq:fp1}
       e_1+e_2  \equiv L  \textnormal{ (mod } \lambda).
    \end{equation}
    \begin{equation}\label{eq:fp2}
        e_1  \equiv e_2  \textnormal{ (mod } \tau).
    \end{equation}
    \begin{equation}\label{eq:fp3}
        e_1  \not\equiv e_2  \textnormal{ (mod } \rho).
    \end{equation}
\end{definition}

\begin{lemma}\label{l:pointstomonomials}
    Let $\sigma \in \{2,3,\rho\}$. Then $\ev_{v,A}(X^{e_1})\cdot_h \ev_{v,A}(X^{e_2})\neq 0$ if and only if $(e_1,e_2)$ is a failure point.
\end{lemma}
\begin{proof}
    This follows directly from Proposition \ref{p:orthogconditions}. 
\end{proof}  

In our study of $\Hull(\C)$, it is essential to understand the orthogonality relations between monomials $X^{e_1}, X^{e_2}$. In particular, we wish to count the number of monomials whose evaluation vectors are not orthogonal under the Hermitian inner-product. By Lemma \ref{l:pointstomonomials}, this is equivalent to counting the number of non-negative integer solutions to Equations (\ref{eq:fp1})-(\ref{eq:fp3}). 
\begin{definition}
    We denote by $\cF_{<k}$ the set of non-negative integer points $(e_1,e_2)$ that satisfy conditions Equations (\ref{eq:fp1})-(\ref{eq:fp3}) such that $\max\{e_1,e_2\}<k$.
\end{definition}
\begin{remark}
    By the symmetry of Equations (\ref{eq:fp1})-(\ref{eq:fp3}), we see that $(e_1,e_2)$ is a failure point $\iff$ $(e_2,e_1)$ is a failure point. Moreover, Equation (\ref{eq:fp3}) excludes points of the form $(x,x)$.  Therefore, to characterize all failure points, it is sufficient to consider only points $(e_1,e_2)$ with $e_1 < e_2$. 

\end{remark}

\begin{lemma}\label{l:countingc}
Let $\sigma\in \{2,3,\rho\}$. If $k\leq \lambda \tau$, or $k\leq  2\lambda \tau$ and $\rho=2$, we have
$$
\Hull(\C)=\langle \ev_{v,A}(X^{i}):i\not \in \pi_1(\cF_{<k} )\rangle_{\F_{q^2}},
$$
where $\pi_1$ is the projection onto the first coordinate. Therefore, $c=\abs{\cF_{<k}}$. Moreover, for any $1\leq k\leq n$ and $\sigma$, we have 
$$
\Hull(\C)\supset \langle \ev_{v,A}(X^{i}):i\not \in \pi_1(\cF_{<k} )\rangle_{\F_{q^2}},
$$
and $c\leq \abs{\cF_{<k}}$. 
\end{lemma}
\begin{proof}
We start by proving that $c=\rk G\cdot (G^q)^t\leq \abs{\cF_{<k}}$. This is because $\abs{\cF_{<k}}$ is the number of nonzero entries of $G\cdot (G^q)^t$ (see Equation (\ref{eq:entries})), which is always greater than or equal to the rank. In what follows, we reason with lattice points $(e_1,e_2)$, but it directly translates to the monomials that belong to $\Hull(\C)$, see Prop. \ref{p:orthogconditions} and Lem. \ref{l:pointstomonomials}. 

If $k\leq \lambda \tau$, note that, given $(e_1,e_2)\in \cF_{<k}$, then $(e_1,e_2+\beta)\not \in \cF_{<k}$ for any $\beta\neq 0$ with $\beta\leq k-1-e_2$. This is because if $(e_1,e_2)$ and $(e_1,e_2+\beta)$ both satisfy Equations (\ref{eq:fp1}) and (\ref{eq:fp3}), then we have $\beta \equiv 0\bmod \lambda$ and $\beta\equiv 0 \bmod \tau$, which implies $\beta \equiv 0 \bmod \lambda \tau$ since $\lambda$ and $\tau$ are coprime. This can only happen for $\beta=0$ or $\beta \geq \lambda \tau>k-1$. A similar argument shows that $(e_1+\beta,e_2)\not \in \cF_{<k}$. By Equation (\ref{eq:entries}), this means that every row and column of $G\cdot (G^q)^t$ has only, at most, one nonzero entry (in terms of monomials, each monomial is not orthogonal to, at most, one other monomial), and then $\rk G\cdot (G^q)^t$ is equal to the number of nonzero entries, which is given by $\abs{\cF_{<k}}$. 

Finally, assume that $k\leq 2\lambda \tau$ and $\rho=2$. If both $(e_1,e_2),(e_1,e_2+\beta)\in \cF_{<k}$, arguing as before, we get $\beta\equiv 0 \bmod \lambda \tau$. By Equation (\ref{eq:fp3}), we also have $e_1\not \equiv e_2\bmod 2$, which implies $e_1-e_2\equiv 1\bmod 2$. Taking into account Equation (\ref{eq:fp3}) again, we also obtain $e_1\not\equiv e_2+\beta\bmod 2$, which translates to $\beta \equiv 0 \bmod 2$. If $\rho=2$, then $\lambda$, $\tau$ and $\rho$ are pairwise coprime (since $\rho/\kappa\geq2$ implies $\kappa_1=\kappa_2=1$), $\lambda \tau$ is odd and then we must have $\beta\equiv 0 \bmod 2\lambda \tau$. Reasoning as in the previous paragraph, we obtain the result. 
\end{proof}

As a consequence of the previous result, the parameters of the EAQECCs obtained from $\C$ and the Hermitian construction \ref{t:hermitian}, when $k\leq \lambda \tau$ (or $k\leq 2\lambda \tau$ and $\rho=2$) and $\sigma\in \{2,3,\sigma\}$ are
\begin{equation}\label{eq:paramQuant}
[[\lambda\tau \sigma,\lambda\tau \sigma-2k+\abs{\cF_{<k}},k+1;\abs{\cF_{<k}}]]_q.
\end{equation}

\begin{lemma}\label{l:eaqmds}
Let $k\leq  \lambda \tau$. Then the Hermitian construction \ref{t:hermitian} applied to $\C$ gives rise to an EAQMDS code. 
\end{lemma}
\begin{proof}
The result follows from Equation (\ref{eq:paramQuant}) (if $\sigma \not \in \{2,3,\rho\}$, a similar expression holds with $c$ instead of $\abs{\mathcal{F}_{<k}}$) by checking that we have equality in the first expression of Theorem \ref{t:singleton}. Note that, since by definition $c\leq \dim \C=k$, the first bound implies the second bound in Theorem \ref{t:singleton} (recall that the minimum distance of the quantum code is $k+1$). Finally, the last bound in Theorem \ref{t:singleton} does not apply, since $k\leq \lambda\tau\leq n/2$. 
\end{proof}

Note that, when $\rho=2$, by Lemma \ref{l:countingc} we can still get the exact value of $c$ with $\abs{\cF_{<k}}$, but the corresponding code may not achieve equality in the last bound of Theorem \ref{t:singleton}. In fact, if we have a quantum code with parameters $[[n,k,d;c]]_q$, constructed using the Hermitian construction, in \cite[Thm. 15]{grasslvariableentanglement} it is shown that
$$
2d\leq n+c-k+2.
$$
Thus, when the last bound in Theorem \ref{t:singleton} is lower than the first, we cannot obtain EAQMDS codes with the Hermitian construction. 

In the next sections, we study generic lattices satisfying conditions similar to Equations (\ref{eq:fp1}) and (\ref{eq:fp2}), which can be used to compute the number of points in $\cF_{<k}$. This in turn gives a description of the monomials in $\Hull(\C)$ via Lemma \ref{l:countingc}, and the parameter $c$ of the corresponding EAQECCs.

\section{Counting points in lattices}\label{s:countinglattices}
We present now a counting problem for a generic lattice whose points satisfy conditions similar to Equations (\ref{eq:fp1})-(\ref{eq:fp3}) in Definition \ref{def_failurepoint}. The problem of counting the points in $\cF_{<k}$ will translate to the problem of counting the points in several lattices of the form we introduce. In what follows, let $B,C>1$ be positive integers, and let $A$ be a non-negative integer. Consider the following equations: 
\begin{equation}\label{eq:genericmodular1}
e_1+e_2  \equiv A  \textnormal{ (mod } B),
\end{equation}
\begin{equation}\label{eq:genericmodular2}
    e_1 \equiv e_2 \textnormal{ (mod }  C).
\end{equation}

\begin{definition}
    A \textbf{lattice point} $(e_1,e_2)$ is a non-negative integer point satisfying the Equations (\ref{eq:genericmodular1}) and (\ref{eq:genericmodular2}), with $e_1<e_2$. The set of all lattice points is denoted $\cL_{A,B,C}$. 
\end{definition}
\begin{remark}\label{r:solutions}
    Every lattice point $(e_1,e_2)$ lies at the intersection point of some pair of lines
    \begin{equation}\label{eq:genericf}
    f(t): y=-x+tB+A,
\end{equation}
\begin{equation}\label{eq:genericg}
g(\varepsilon) : y=x+\varepsilon C,
\end{equation}
for a unique value of $t,\varepsilon$, hence our use of the term lattice. We will use the notation $(e_1,e_2)\in f(t)\cap g(\varepsilon)$. For a given $t,\varepsilon$, the intersection point is a lattice point if and only if 
\begin{equation}
    tB+A-\varepsilon C
\end{equation}
is an even non-negative integer. There is always some values for $t, \varepsilon$ such that this is true, except with $B,C$ even and $A$ odd, in which case $\cL_{A,B,C}$ is empty. For convenience, we will assume that $\cL_{A,B,C}\neq \emptyset$ except when otherwise specified. 
\end{remark}

\begin{lemma}\label{l:e>0}
    Let $(e_1,e_2)\in f(t)\cap g(\varepsilon)$ be a lattice point. Then $\varepsilon\ge1$. 
\end{lemma}
\begin{proof}
    Since $(e_1,e_2)\in g(\varepsilon)$ we have that $e_2=e_1+\varepsilon C$. Since $e_1<e_2$ it follows that $\varepsilon >0$. 
\end{proof}

\begin{remark}\label{r:moves}
It is useful to consider ``moves'' on the lattice, of which there are two types: 
\[
(x,y)\mapsto (x,y)\pm (B/2,B/2),
\]
\[
(x,y)\mapsto(x,y)\pm(-C/2,C/2),
\]
along with their inverses. For a given lattice point $(e_1,e_2) \in f(t')\cap g(\varepsilon')$, the first move brings us to $f(t'\pm1)\cap g(\varepsilon')$, and the second move brings us to $f(t')\cap g(\varepsilon'\pm1)$. These may or may not be integer points, depending on the parities of $B$ and $C$, but every integer point can be reached from the others by some combination of these moves. Note that even if one of these moves may not produce an integer point starting from another, a combination of them might, e.g., the move 
$$
(x,y)\mapsto (x,y)\pm \left(\frac{B-C}{2},\frac{B+C}{2}\right)
$$
maps integer points to integer points when both $B$, $C$ are odd, even though a single one of the moves presented above does not.
\end{remark}
\begin{lemma}\label{l:pointtopoint}
    Let $(e_1,e_2)$ and $(e_1',e_2')$ be two lattice points located at $f(t)\cap g(\varepsilon)$ and $f(t')\cap g(\varepsilon')$ respectively. Then there exist unique integers $i,j$ such that 
    \[
    (e_1',e_2')=(e_1,e_2)+i(B/2,B/2)+j(-C/2,C/2).
    \]
\end{lemma}
\begin{proof}
    Starting from $f(t)\cap g(\varepsilon)$, the move $+i(B/2,B/2)+j(-C/2,C/2)$ takes us to $f(t+i)\cap g(\varepsilon+j)$. Taking $i=(t'-t), j=(\varepsilon'-\varepsilon)$ will yield the desired result, and it is clear that $i,j$ are unique. 
\end{proof}
Our goal is to calculate $|\cL_{A,B,C}|$ for lattice points in a certain range. Our general strategy will be to find a suitable starting point on the lattice, and to count how many moves we can make while remaining inside $\cL_{A,B,C}$.

\begin{definition}\label{d:firstlatticepoint}
Given the lattice $\cL_{A,B,C}$, we define the \textbf{first lattice point} $(D_1,D_2)$ to be the point such that for any $(e_1,e_2) \in \cL_{A,B,C}$, either $D_2 \leq e_2$ or $D_2=e_2$ and $D_1<e_1$. We will denote by $t^*,\varepsilon^*$ the values such that $(D_1,D_2) \in f(t^*)\cap g(\varepsilon^*)$. 
\end{definition}

\begin{remark}
The first lattice point is the minimal element of $\cL_{A,B,C}$ with respect to the colexicographic ordering (or reflected lexicographic ordering). We will often implicitly refer to this ordering, saying that a point is ``smaller'' than another or ``minimal'' within a set. 
\end{remark}

\begin{lemma}[\hspace{1pt}{\cite[Lem. 5.1]{campionQMDS}}]\label{l:square}
Suppose $B > C$. Let $Q,Q'$ be positive integers with $Q'>Q$. Consider the four lines: 
    \begin{equation*}
        \begin{split}
            & L_1: y=-x+QB +L.\\
            & L_2: y=-x+Q'B +L.\\
            & L_1': y=x + B.\\
            & L_2': y=x + 2B.\\
        \end{split}
    \end{equation*}
    Consider also the four points: 
       \begin{equation*}
        P_{ij} := (\alpha_{ij},\beta_{ij}):= L_i\cap L_j'
    \end{equation*}
    Then $\text{max}\{\beta_{11},\beta_{12}\} < \text{min}\{\beta_{21},\beta_{22}\}$
\end{lemma}

\begin{lemma}\label{l:1}
If there is a lattice point on $f(t)$, then there is a lattice point at $f(t)\cap g(\varepsilon)$ with $\varepsilon\in \{1,2\}$. Moreover, if $C$ is even, then there is a lattice point at $f(t)\cap g(1)$. 
\end{lemma}
\begin{proof}
Suppose the lattice point on $f(t)$ is located at $f(t)\cap g(\varepsilon')$. Applying the move $(C,-C)$ will bring us to the lattice point $f(t)\cap g(\varepsilon'-2)$, and we can repeat until we are at $f(t)\cap g(\varepsilon)$ with $\varepsilon \in \{1,2\}$. Moreover, if $C$ is even then the move $(C/2,-C/2)$ will also bring us from lattice points to lattice points, ensuring that $f(t)\cap g(1)$ is a lattice point. 
\end{proof}

\begin{lemma}\label{l:2}
Let $(D_1,D_2)$ be on the line $g(\varepsilon^*)$. Then $\varepsilon^*\in \{1,2\}$. Moreover, if $C$ is even then $\varepsilon^*=1$.
\end{lemma}
\begin{proof}
    This follows from Lemma \ref{l:1} and the minimality of $D_2$. 
\end{proof}
\begin{lemma}\label{l:3}
Let $(D_1,D_2)$ lie on the line $f(t^*)$. Then for any $t<t^*$,   $\cL_{A,B,C}\cap f(t)=\emptyset$. 
\end{lemma}
\begin{proof}
We argue by contradiction. Suppose that there is a lattice point $(e_1,e_2)\in \cL_{A,B,C}$ on the line $f(t)$ with $t<t^*$. By Lemma \ref{l:1}, there is a point on the line $f(t)$ with 
$$
e_2-e_1=\varepsilon'C,
$$
with $\varepsilon'\in \{1,2\}$. By Lemma \ref{l:2}, $(D_1,D_2)$ has $D_2-D_1=\varepsilon C$, with $\varepsilon\in \{1,2\}$. We now consider a number of cases. 
\begin{itemize}
    \item If $\varepsilon=2$, then $e_1+e_2<D_1+D_2$ (since $t<t^*$), and $e_2-e_1\leq D_2-D_1$ (because $\varepsilon'\leq \varepsilon$). Adding the two conditions, this implies $e_2<D_2$, a contradiction. 
    \item If $\varepsilon=1$, $\varepsilon'=2$ and $B> C$, then we apply Lemma \ref{l:square} to find that $e_2<D_2$, a contradiction.
    \item If $\varepsilon=1$, $\varepsilon'=2$ and $B<C$, then using Lemma \ref{l:pointtopoint}, we can write
$$
(e_1,e_2)=(D_1,D_2)+(-C/2,C/2)-i(B/2,B/2),
$$
with $i=t^*-t'>0$. 
If $B$ and $C$ have the same parity, then consider the point 
\[
(x,y)=(e_1,e_2)+\left(\frac{C-B}{2}, \frac{-C-B}{2}\right)
\]
This is an integer point, since $B$ and $C$ have the same parity. Moreover, since $B<C$, we have that $x\ge 0$. Also, $y-x=C$. So $(x,y)$ is in the lattice, and $y=D_2-(i+1)(B/2,B/2)<y$, contradicting the minimality of $D_2$.

If $C$ is odd and $B$ is even, then $(e_1,e_2)$ cannot be an integer point. Finally, if $C$ is even, and $B$ is odd, then by Lemma \ref{l:2}, $\varepsilon^*=2$, which contradicts our earlier assumption. 
\end{itemize}

\end{proof}
\begin{lemma}\label{l:4}
    Let $t'$ be the least integer such that $f(t')$ contains a lattice point. Then the first lattice point is located at $f(t')\cap g(\varepsilon')$, where $\varepsilon'=\min\{\varepsilon:f(t')\cap g(\varepsilon) \in \cL_{A,B,C}$.
\end{lemma}
\begin{proof}
    Let $(D_1,D_2) \in f(t^*)\cap g(\varepsilon^*)$ be the first lattice point, and denote by $(x,y)$ the lattice point at $f(t')\cap g(\varepsilon')$. By definition we have $t'\leq t^*$, and by Lemma \ref{l:3} we have that $t^*\leq t'$; thus $t'=t^*$. We have that $\varepsilon' \leq \varepsilon^*$ by definition, and we must have equality, otherwise we would contradict the minimality of $D_2$. Therefore  $(x,y)=(D_1,D_2)$. 
\end{proof}
This lemma gives us a useful characterization of the first lattice point. In order to count the number of lattices points correctly, we want to make sure that we only have to make moves of type $+(B/2,B/2)$ and $+(-C/2,C/2)$ from the first lattice point, and not their inverses. We will call such moves ``positive moves''. This happens only in some cases: 
\begin{proposition}[Positive moves]\label{p:posmoves}
   Let $(D_1,D_2) \in f(t^*)\cap g(\varepsilon^*)$ be the first lattice point on $\cL_{A,B,C}$. Then every other point $(e_1,e_2)$ on the lattice can be written uniquely as 
    \[
    (e_1,e_2) = (D_1,D_2)+i(B/2,B/2)+j(-C/2,C/2)
    \]
    with $i\ge 0$. Moreover, if $(D_1,D_2) \in g(1)$ then $j\ge0$. 
\end{proposition}
\begin{proof}
    Let $(e_1,e_2) \in f(t)\cap g(\varepsilon)$. It follows from Remark \ref{r:moves} and a simple calculation that $(e_1,e_2)$ can be written as:
    \[
    (e_1,e_2)=(D_1,D_2)+(t-t^*)(B/2,B/2)+(\varepsilon-\varepsilon^*)(-C/2,C/2).
    \]
    Uniqueness is trivial. By assumption, $(t-t^*)\ge 0$, and if $\varepsilon^*=1$ then it follows from Lemma \ref{l:e>0} that $(\varepsilon-\varepsilon^*)\ge 0$. 
\end{proof}
Next, we consider the problem that not all moves with positive coefficients are valid. For example, if $B$ is odd, then the move $+(B/2,B/2)$ will take us from an integer point to a non-integer point. To deal with this, we will divide the lattice into two sub-lattices, where every move with positive coefficients is valid. This will allow us to easily count the number of lattice points. 
\begin{definition}\label{d:partition}
    Given the lattice $\cL_{A,B,C}$ and the first lattice point $(D_1,D_2)\in f(t^*)\cap g(\varepsilon^*)$, we partition the lattice into two subsets: 
    \[
    \cL_{A,B,C}^1:=\{(e_1,e_2)\in \cL_{A,B,C}: (e_1,e_2)\in f(t) \text{ with } t\equiv t^* (\text{mod } 2)\},
    \]
    \[
        \cL_{A,B,C}^2:=\{(e_1,e_2)\in \cL_{A,B,C}: (e_1,e_2)\in f(t) \text{ with } t\not\equiv t^* (\text{mod } 2)\}.
    \]
\end{definition}

\begin{lemma}\label{l:sublattices}
    Using the notation from the beginning of this section, the set $\cL_{A,B,C}^1$ is a lattice $\cL_{(A+t^*B), 2B, C}$ and $\cL_{A,B,C}^2$ is a lattice $\cL_{(A+(t^*+1)B), 2B, C}$.
\end{lemma}
\begin{proof}
    The points of $\cL_{A,B,C}^1$ (and similarly for $\cL_{A,B,C}^2$) are given by intersection points of the parametric families of lines:
\[
f_1(t): y=-x+t(2B)+(t^*B+A), ~~~~g_1(\varepsilon) : y=x+\varepsilon C
\]
which corresponds precisely to the solutions of the modular equations:
\begin{enumerate}
    \item $e_1+e_2  \equiv A+t^*B $ (mod $2B$),
    \item $e_1 \equiv e_2$ (mod $C$).
\end{enumerate}
\end{proof}
\begin{remark}
    The notation of Lemma \ref{l:sublattices} is not particularly useful,  and we will stick to writing $\cL_{A,B,C}^1$ and $\cL_{A,B,C}^2$. The main point is that these sets are lattices in their own right, and so we can apply the definitions and results of the generic lattices $\cL_{A,B,C}$ to these lattices also. 
\end{remark}
Next, we compute the first lattice point on each of the sub-lattices, and show how we can count the points on each using only positive moves. 

\begin{lemma}\label{l:firstpointL1}
    Let $(D_1,D_2) \in f(t^*)\cap g(\varepsilon^*)$ be the first lattice point on $\cL_{A,B,C}$. Then $(D_1,D_2)$ is also the first lattice point on $\cL_{A,B,C}^1$. 
\end{lemma}
\begin{proof}
    By definition we have $(D_1,D_2)\in \cL_{A,B,C}^1$. Since $\cL_{A,B,C}^1 \subseteq \cL_{A,B,C}$, it follows by minimality that $(D_1,D_2)$ is also the first lattice point for $\cL_{A,B,C}^1$. 
\end{proof}

\begin{proposition}[Positive Moves on Lattice 1]\label{p:posmoves1}
        Let $(D_1,D_2) \in f(t^*)\cap g(\varepsilon^*)$ be the first lattice point on $\cL_{A,B,C}$.  Then every point in $\cL_{A,B,C}^1$ can be written uniquely as
    \begin{equation}\label{eq:L1moves}
            (e_1,e_2) = (D_1,D_2)+i(B,B)+j(-C/2,C/2)
    \end{equation}
with $i,j\ge 0$.
\end{proposition}
\begin{proof}
    By Lemmas \ref{l:sublattices} and \ref{l:firstpointL1}, $(D_1,D_2)$ is the first lattice point of the lattice $\cL_{A,B,C}^1$. It follows from Proposition \ref{p:posmoves} that every point of $\cL_{A,B,C}^1$ can be written uniquely as in Equation (\ref{eq:L1moves}) with $i\ge 0$. 
    
    It is sufficient to show that if $\varepsilon^*=2$, then $\cL_{A,B,C}^1\cap g(1)=\emptyset$. Suppose there is a point in $\cL_{A,B,C}^1\cap g(1)$, and suppose that the point is at $f(t')\cap g(1)$ for some $t'$. By definition, $t'-t^*$ is even, and applying the move $\frac{(t'-t^*)}{2}(-B,-B)$ would give a lattice point at $f(t^*)\cap g(1)$, contradicting the minimality of $D_2$. 
\end{proof}

To find the first lattice point on $\cL_{A,B,C}^2$, it is easier to analyze some specific cases. 

\begin{lemma}\label{l:firstpointL2}
    Suppose that $\cL_{A,B,C}\neq \emptyset$. Then each of the following is true: 
    \begin{enumerate}
        \item We have $\cL_{A,B,C}^2=\emptyset \iff$ $C$ is even and $B$ is odd. 
        \item If $B$ is even then the first lattice point of $\cL_{A,B,C}^2$ is $(D_1,D_2)+(B/2,B/2)$. 
        \item If both $B$ and $C$ are odd, then first lattice point of $\cL_{A,B,C}^2$ is 
        \begin{itemize}
            \item $(D_1,D_2)+((B+C)/2,(B-C)/2)$ if  $\varepsilon^*=2$,
            \item $(D_1,D_2)+((B-C)/2,(B+C)/2)+l(B,B)$ where $l=\lceil(\frac{C-B}{2}-D_1)/B\rceil $, if $\varepsilon^*=1$. 
        \end{itemize} 
    \end{enumerate}
\end{lemma}
\begin{proof}
    (1) Consider the quantity $tB+A-\varepsilon C$, and recall that we have lattice points at $f(t)\cap g(\varepsilon)\iff$ the quantity is a non-negative even integer. If $C$ is even and $B$ is odd, then all solutions in $t$ must have the same parity as $t^*$, or else the quantity will not be even. Hence $\cL_{A,B,C}^2=\emptyset$. Conversely, either $B$ is even, in which case the parity of $t^*$ is irrelevant, or $C$ is odd, in which case we can adjust $\varepsilon$ to fix the parity for a given $t$.
    
    (2) The given point is clearly an integer point, and lies on $f(t^*+1)\cap g(\varepsilon^*)$. Since $t^*$ is minimal for $\cL_{A,B,C}$, $t^*+1$ is minimal for $\cL_{A,B,C}^2$, and thus the first lattice point of $\cL_{A,B,C}^2$ lies on $f(t^*+1)$. Moreover, $\varepsilon^*$ must be minimal on $f(t^*+1)$; otherwise, if there is a lattice point at $f(t^*+1)\cap g(\varepsilon') $ with $\varepsilon' < \varepsilon^*$, then since $B$ is even there is a lattice point at $f(t^*)\cap g(\varepsilon')$, contradicting the minimality in Lemma \ref{l:4}.  Thus, applying Lemma \ref{l:4} to $\cL_{A,B,C}^2$, the given point is the first lattice point on $\cL_{A,B,C}^2$. 

    (3) In the case of $\varepsilon^*=2$, the given point lies at $f(t^*+1)\cap g(1)$. By Lemma $\ref{l:e>0}$ and minimality of $t^*$, it follows again from Lemma \ref{l:4} that the given point is the first lattice point of $\cL_{A,B,C}^2$.  
    
    In the case of $\varepsilon^*=1$, it follows from parities that $\cL_{A,B,C}^2\cap g(1)=\emptyset$. Therefore, the first lattice point lies on $g(2)$, which we can see the given point lies on. Moreover, by Lemma \ref{l:4}. the first lattice point will lie on the smallest $t$ for which $f(t)\cap g(2)$ is a non-negative integer point. The choice of $l$ for the given point ensures minimality of $t$; therefore, this is the first lattice point for $\cL_{A,B,C}^2$. 
\end{proof}
\begin{proposition}[Positive Moves on Lattice 2]\label{p:posmoves2}
        Let $(D_1',D_2') \in f(t)\cap g(\varepsilon)$ be the first lattice point on $\cL_{A,B,C}^2$.  Then every point in $(e_1,e_2)\in\cL_{A,B,C}^2$ can be written uniquely as 
    \begin{equation}\label{eq:L2moves}
            (e_1,e_2) = (D_1',D_2')+i(B,B)+j(-C/2,C/2)
    \end{equation}

    with $i,j\ge 0$.
\end{proposition}
\begin{proof}
    Let $(e_1,e_2) \in f(t')\cap g(\varepsilon')$. Existence follows from Proposition \ref{p:posmoves}, considering the lattice $\cL_{A,B,C}^2$ with first lattice point $(D_1',D_2')$. From the same proposition we have that $i=(t'-t)\ge0, j=(\varepsilon'-\varepsilon)$. To show that $j\ge0$, it suffices to show that if $\varepsilon=2$ then $\cL_{A,B,C}^2\cap g(1)=\emptyset$. By Lemma \ref{l:firstpointL2}, if $\varepsilon=2$ then we are in one of the following cases:
    \begin{itemize}
        \item First consider if $B$ is even. The first lattice point of $\cL_{A,B,C}$ lies at $f(t^*)\cap g(\varepsilon^*)$, and by Lemma \ref{l:firstpointL2}, $(D_1',D_2') \in f(t^*+1)\cap g(\varepsilon)$. So if $\varepsilon=2$, then $\varepsilon^*=2$; it must then follow that $\cL_{A,B,C}\cap g(1)=\emptyset$. Otherwise, since $B$ is even we could apply $(-B/2,-B/2)$ to get a lattice point at $f(t^*)\cap g(1)$, contradicting the minimality of $(D_1,D_2)$ in $\cL_{A,B,C}$.
        \item The only other case is with $B,C$ odd and $\varepsilon^*=2$. As stated in Lemma \ref{l:firstpointL2}, this implies that $\cL_{A,B,C}^2\cap g(1)=\emptyset$, so we are done. 
    \end{itemize}
\end{proof}

We now have the requisite results to derive a formula for counting the number of points on a lattice $\cL_{A,B,C}$ within a specified range.
\begin{definition}
    Let $k\ge 0$ be an integer. Given a lattice $\cL_{A,B,C}$, we define 
    \[
    \cL_{A,B,C,<k}:=\{(e_1,e_2)\in \cL_{A,B,C}\mid 0\leq e_1<e_2<k\} 
    \]
\end{definition}

\begin{proposition}\label{p:numberofpoints}
    Suppose that $\cL_{A,B,C,<k}\neq \emptyset$. Let $(D_1,D_2) $ and $(D_1',D_2')$ be the first lattice points of $\cL_{A,B,C}^1$ and $ \cL_{A,B,C}^2$, which we assume to be non-empty. Then if $C$ is odd we have that 
    \begin{equation}
    |\cL_{A,B,C,<k}^1|=\sum_{i=0}^{\left\lceil (k-D_2-B)/B\right\rceil}  \left(\min\left\{ \left\lfloor\frac{D_1+iB}{C}\right\rfloor,  \left\lceil\frac{k-D_2-iB}{C}-1\right\rceil\right\}+1\right). 
    \end{equation}
The formula for $|\cL_{A,B,C,<k}^2|$ is the same, but with $(D_1',D_2')$ in place of $(D_1,D_2)$. If $C$ is even, then the formulas are the same but with $C/2$ in place of $C$. It then follows that 
\begin{equation}\label{eq:sumsublattice}
   |\cL_{A,B,C,<k}|= |\cL_{A,B,C,<k}^1|+ |\cL_{A,B,C,<k}^2|
\end{equation}

\end{proposition}
\begin{proof}
    It follow directly from Definition \ref{d:partition} that $\cL_{A,B,C,<k}$ is the disjoint union of the sets $\cL_{A,B,C,<k}^1$ and $\cL_{A,B,C,<k}^2$, which easily proves Equation (\ref{eq:sumsublattice}).
    
    First suppose that $C$ is even. By Lemma \ref{p:posmoves1}, every point of $|\cL_{A,B,C,<k}^1|$ can be written uniquely as  
    \[
(e_1,e_2) = (D_1,D_2)+i(B,B)+j(-C/2,C/2)
    \]
    By uniqueness, counting the number of lattice points is equivalent to counting the possible values for $i,j$ for which the given point is a valid lattice point.
    For $j=0$, the restriction that $e_2<k$ implies that $0 \leq i <(k-D_2)/B$, which gives us the limit in the summation. For each given $i$, the restriction that $e_1\ge0$ implies that $0 \leq j \leq 2(D_1+iB)/C$, and the restriction $e_2<k$ implies that $0 \leq j < 2(k-D_2-iB)/C$;  the summand is therefore
    
    $$\min\left\{ \left\lfloor\frac{2(D_1+iB)}{C}\right\rfloor, \left\lceil\frac{2(k-D_2-iB)}{C}-1\right\rceil\right\}+1.$$

    If $C$ is odd, we only consider the moves $i(B,B)+j(-C,C)$, and a similar analysis shows that the summand must be 
     $$\min\left\{ \left\lfloor\frac{D_1+iB}{C}\right\rfloor, \left\lceil\frac{k-D_2-iB}{C}-1\right\rceil\right\}+1.$$

    The proof for $|\cL_{A,B,C,<k}^2|$ is identical.

\end{proof}

\begin{remark}\label{r:whenempty}
     The assumption that the lattices be non-empty is no restriction on our ability to calculate $|\cL_{A,B,C,<k}|$. We summarize the situation in the following:
     \begin{itemize}
         \item If $k\leq D_2 $ then by minimality of $D_2$ we have that $|\cL_{A,B,C,<k}|=0$. 
         \item If $D_2 < k \leq D_2' $ then $0 \neq |\cL_{A,B,C,<k}|=|\cL_{A,B,C,<k}^1|$ since $\cL_{A,B,C,<k}^2=\emptyset$.
         \item If $D_2'<k$ then $ |\cL_{A,B,C,<k}|= |\cL_{A,B,C,<k}^1|+ |\cL_{A,B,C,<k}^2|$, with $|\cL_{A,B,C,<k}^2|=0 \iff C$ is even and $B$ is odd. 
         \end{itemize}

\end{remark}
\section[The first lattice point of T]{The first lattice point of $\cT$}\label{s:laticeT}
We return now to the problem of calculating $|\cF_{<k}|$. While this lattice is similar to those studied in the previous section, it is in fact not of the same type. In order to use those results, we will decompose the lattice $\cF_{<k}$ into those of the correct type. 

Recall the notation of Section 1; in particular, we will recall the conditions of a failure point $(e_1,e_2)$:  
        \begin{enumerate}
    \item $e_1+e_2  \equiv L $ (mod $\lambda$).
    \item $e_1 \equiv e_2$ (mod $\tau$).
    \item $e_1\not\equiv e_2$ (mod $\rho$).
\end{enumerate}

\begin{definition}\label{d:lattices}
    Let $k\ge 0$ be an integer. We define the lattices: 

\begin{equation*}
    \begin{split}
         \cB_{<k}&:=\{(e_1,e_2)\mid 0\leq e_1,e_2< k\},\\
\cF^+&:=\{ (e_1,e_2)\mid  e_1 < e_2 \textnormal{ and $(e_1,e_2)$ satisfy conditions (1), (2) and (3)}\},\\
\cT&:=\{ (e_1,e_2)\mid  e_1 < e_2 \textnormal{ and $(e_1,e_2)$ satisfy conditions (1) and (2)}\},\\
\cP&:=\{ (e_1,e_2)\mid  e_1 < e_2 \textnormal{ and $(e_1,e_2)$ satisfy conditions (1) and (2), but not (3)}\}.\\
    \end{split}
\end{equation*}
\end{definition}
We use the notation $\cF_{<k}^+:=\cF^+\cap\cB_{<k}$, similarly for $ \cT_{<k}$ and $\cP_{<k}$. Note that by the paragraph following Lemma \ref{l:pointstomonomials} we have that $ |\cF_{<k}|=2|\cF_{<k}^+|$, where $\cF_{<k}$ is as defined in Section \ref{s:orthog}. It is also clear from the definition that $\cP \subset \cT$ and $\cF^+=\cT\setminus \cP$.

\begin{remark}\label{r:TandPlattices}
    The lattices $\cT$ and $ \cP$ are precisely of the type studied in Section \ref{s:countinglattices}. Explicitly, the lattice $\cT$ can be written as $\cT=\cL_{L,\lambda,\tau}$ and is given by the equations 
    \begin{enumerate}
    \item $e_1+e_2  \equiv L$ (mod $\lambda$).
    \item $e_1 \equiv e_2$ (mod $\tau$).
\end{enumerate}
Similarly the lattice $\cP=\cL_{L,\lambda,\tau\rho/\kappa_1}$ is given by the equations 
    \begin{enumerate}
    \item $e_1+e_2  \equiv L$ (mod $\lambda$).
    \item $e_1 \equiv e_2$ (mod $\tau\rho/\kappa_2$).
\end{enumerate}
In this section, we will examine the lattice $\cT$ and use the results from Section \ref{s:countinglattices}, with $A=L,B=\lambda, C=\tau$. We will also use $f,g$ to refer to the parametric lines 
\[
f(t): y=-x+t\lambda+L, ~~~~g(\varepsilon) : y=x+\varepsilon \tau.
\]
 
\end{remark}

\begin{definition}
    We will denote by $(T_1,T_2)$  and $(P_1,P_2)$ the first lattice points of $\cT$ and $\cP$ respectively. 
\end{definition}

Our next task will be to compute explicit formulas for the values of $(T_1,T_2)$  and $(P_1,P_2)$. First note that while the lattice $\cF$ is not of the same type as those on Section \ref{s:countinglattices}, the notion of the ``first lattice point'' as in Definition \ref{d:firstlatticepoint} is still well defined. In \cite{campionQMDS}, the authors found explicit formulas for the first lattice point of $\cF^+$, which we will denote $(F_1,F_2)$ and write down here:

    \begin{table}[h!]
        \centering
\begin{tabular}{|| c|c| c ||} 
 \hline
Conditions & Value of $L$ &$(F_1,F_2)$\\  
 \hline\hline
$\lambda$ even 
&$2\tau-2$
 & $\left(\frac{\lambda-2}{2},\frac{\lambda+4\tau-2}{2}\right)$\\ 
 \hline
\specialcell{$\lambda$ odd, and at least one of the following\\
 conditions holds: $\lambda<\tau$,  $\tau$ even, $\rho=2$}
&$\tau-2$
 & $\left(\lambda-1,\lambda+\tau-1\right)$    \\
 \hline
$ \lambda $ odd, $\lambda >\tau, \tau $ odd, $\rho \neq 2$ 
&$2\tau-2$
 & $\left(\frac{\lambda+\tau-2}{2},\frac{\lambda+3\tau-2}{2}\right)$  \\
 \hline
\end{tabular}
        \caption{Values of $L$ and $(F_1,F_2)$}
        \label{tab:valueL}
    \end{table}

In almost all cases, the first lattice point of $\cF^+$ is the same as $(T_1,T_2)$, which we now make precise.
\begin{proposition}[Values of $(T_1,T_2)$]
    If we have $\lambda$ odd, $\tau$ odd, $\rho=2$ and $\lambda \ge \tau+2$ then the first lattice point of $\cT$ is   $\left(\frac{\lambda-\tau-2}{2},\frac{\lambda+3\tau-2}{2}\right)$.
    Otherwise, $(T_1,T_2)$ is the same as the first lattice point of $\cF^+$ as given in the table above. 
\end{proposition}
\begin{proof}
    We will address each case in the table above separately. In this proof, we will refer to the points of $\cT$ as lattice points. It follows from Lemma \ref{l:2} that $(T_1,T_2)$ is the minimal lattice point of the set $\cT \cap (g(1)\cup g(2))$, and that $(F_1,F_2)$ is the minimal lattice point of the set $\cF \cap (g(1)\cup g(2))$. Since $\cF^+=\cT\setminus \cP$, we just need to investigate whether subtracting $\cP$ makes a difference. It follows from the definition that $\cP \cap g(1)=\emptyset$, else we would have $\rho \mid \tau$, contradicting the assumption that $\sigma\ge2$. Thus, the only set to consider is $\cP \cap g(2)$. 

    If we are in the first case from Table \ref{tab:valueL} then $(F_1,F_2)\in g(2)$. Since $(F_1,F_2) \in \cT\setminus\cP$, it follows that $\rho \nmid 2\tau$, hence $\cP \cap g(2)=\emptyset$. It follows from the previous discussion that $\cT \cap (g(1)\cup g(2)) = \cF\cap (g(1)\cup g(2))$ and $(F_1,F_2)=(T_1,T_2)$. 

    Next we consider the second case in Table \ref{tab:valueL}, where $(F_1,F_2) \in g(1)$.
    \begin{itemize}
        \item If $\tau$ is even then $(T_1,T_2) \in g(1)$ by Lemma \ref{l:2}. Since $\cP\cap g(1)=\emptyset$ it follows that $(F_1,F_2)=(T_1,T_2)$.
        \item If $\lambda < \tau$, we claim that $(T_1,T_2)\in g(1)$. Otherwise, we would have that $(T_1,T_2)=(F_1,F_2)+i(\lambda/2,\lambda/2)+(-\tau/2,\tau/2)$ for some $i$. Since $F_1=\lambda-1$, the restriction $T_1\ge0$ forces $i \ge0$. However, this would imply that $F_2<T_2$, contradicting the minimality of $(T_1,T_2)$.  
        \item Finally suppose that $\rho=2$. Then $(F_1,F_2)$ is the minimum point on the line $\cT \cap g(1)$; the question is whether there is a smaller point on $\cT \cap g(2)$. Suppose there is; as in the previous paragraph, we have that $(T_1,T_2)=(\lambda-1,\lambda+\tau-1)+i(\lambda/2,\lambda/2)+(-\tau/2,\tau/2)$. In order to satisfy $T_2 < F_2$ it must be that $i=-1$; the constraint $T_1 \ge0 $ then forces $\lambda \ge \tau+2$. So if $\lambda < \tau+2$, then there is no such point, and thus $(T_1,T_2)=(F_1,F_2)$.
        If $\lambda \ge \tau +2$, then the point $\left(\frac{\lambda-\tau-2}{2},\frac{\lambda+3\tau-2}{2}\right)$ is $(T_1,T_2)$. This is an integer point due to the assumed parities and it satisfies $T_2<F_2$ and $T_1\ge 0$. Moreover, $T_1 -i\lambda/2<0$ for any $i \ge 0$, so the point is minimal on the line $g(2)$. 
        \end{itemize}

Finally, let us consider the third case in Table \ref{tab:valueL}. We observe that $(F_1,F_2) =\left(\frac{\lambda+\tau-2}{2},\frac{\lambda+3\tau-2}{2}\right)\in g(1)$, and it must be the minimal lattice points on that line. All lattice point on $g(2)$ can be written as $\left(\frac{\lambda+\tau-2}{2},\frac{\lambda+3\tau-2}{2}\right)+i(\lambda/2,\lambda/2)+(-\tau/2,\tau/2)$. The restriction that the first coordinate must be non-negative means that $i\ge 0$, so the second coordinate is $>F_2$. Therefore $(T_1,T_2)=(F_1,F_2)$. 
\end{proof}
\begin{remark}\label{r:valuesT1prime}
 By Lemma \ref{l:firstpointL1}, the first point on the sub-lattice $\cT^1$ is the same as $(T_1,T_2)$, and we can calculate the first point on the sub-lattice $\cT^2$ using Lemma \ref{l:firstpointL2}.
\end{remark} 

\begin{theorem}[Values of $(T_1,T_2)$]\label{t:valuesT1T2}
    Let $\cT$ be as in Definition \ref{d:lattices}. Then the first lattice point of $\cT$ is given by Table \ref{tab:valueT1}, with $l=\left\lceil(\frac{\tau-\lambda}{2}-T_1)/\lambda\right \rceil$.
\end{theorem}
       \begin{table}[h!]
        \centering
\begin{tabular}{|| c |c |c||} 
 \hline
 Conditions & $(T_1,T_2)$ \\  
 \hline\hline
 $\lambda$ even
 &$\left(\frac{\lambda-2}{2},\frac{\lambda+4\tau-2}{2}\right)$ 
 \\ 

 \hline
\specialcell{$\lambda$ odd, $\tau$ even }
 & $\left(\lambda-1,\lambda+\tau-1\right)$ \\
 \hline
 \specialcell{$\lambda$ odd, $\tau$ odd, \\$\rho=2$, $\lambda < \tau+2$}
 & $\left(\lambda-1,\lambda+\tau-1\right)$ \\
  \hline
 \specialcell{$\lambda$ odd, $\tau$ odd, \\$\rho=2$, $\lambda \ge \tau+2$}
 &  $\left(\frac{\lambda-\tau-2}{2},\frac{\lambda+3\tau-2}{2}\right)$\\
 \hline
  \specialcell{$\lambda$ odd, $\tau$ odd,\\$\rho\neq2$, $\lambda<\tau$ }
 & $\left(\lambda-1,\lambda+\tau-1\right)$\\

 \hline
\specialcell{$ \lambda $ odd, $\lambda >\tau,$ \\ $\tau $ odd, $\rho \neq 2$}
 & $\left(\frac{\lambda+\tau-2}{2},\frac{\lambda+3\tau-2}{2}\right)$ \\
 \hline
\end{tabular}
        \caption{Values of $(T_1,T_2)$}
        \label{tab:valueT1}
    \end{table}

\section[]{The First Lattice point of $\cP$}\label{s:latticeP}

Next we will calculate the values of $(P_1,P_2)$. There are many cases, but all can be described in a systematic way. The points of $ (x,y) \in \cP$ lie at the intersection points of the parametric lines:
\[
f(t): y=-x+t\lambda+L, ~~~~g(\varepsilon) : y=x+\varepsilon \pi. 
\]
where $\pi := \tau\rho/\kappa_2 = \textrm{lcm}(\tau,\rho)$. For the values above, we will write $(x,y) \in f(t) \cap g(\varepsilon)$. For a given $t,\varepsilon$, we can solve the equations to find that $x= (t\lambda+L-\varepsilon\pi)/2.$
Let 
$$
\beta(\varepsilon)= \varepsilon\pi-L.
$$
Thus, the point $(x,y) \in f(t)\cap g(\varepsilon)$ is an integer point if and only if the quantity
$t\lambda-\beta(\varepsilon)$
is an even-non-negative integer. For a fixed $\varepsilon$, this implies that $t \ge \frac{\beta(\varepsilon)}{\lambda}$. By Lemma \ref{l:4}, the first lattice point is characterized by the smallest value of $t$ for which there is a solution to above equation. Therefore it follows that the value of $t$ for the first lattice point is the smallest integer greater than $\frac{\beta(\varepsilon)}{\lambda}$, with the correct parity such that the quantity $t\lambda-\beta(\varepsilon)$ is even. 

This analysis works, provided that we already know the value of $\varepsilon$ for the first lattice point. Writing $(P_1,P_2) \in f(t^*)\cap g(\varepsilon^*)$, by Lemma \ref{l:2} we have that $\varepsilon^* \in \{1,2\}$. In some cases, we can determine the value of $\varepsilon^*$ from the parities the parameters; in other cases, we will have to compare the smallest lattice point on each of the lines $g(1), g(2)$. Let us proceed with a case-by-case analysis. 
\begin{itemize}
    \item Case 1 ($\lambda$ even, $\rho$ odd): From the table in Section \ref{s:prelims}, we see that $L=2\tau-2$, which is even. Since $\lambda$ is even, $\tau$ must be odd, so we have that $\pi$ is odd. We require $\beta(\varepsilon^*)$ to be even, which forces $\varepsilon^*=2$. The parity of $t$ is not relevant, since it only appears multiplied by $\lambda$. Therefore, $t^*= \left\lceil\frac{\beta(\varepsilon)}{\lambda}\right\rceil$. 
    \item Case 2 ($\lambda$ even, $\rho$ even): Since $\pi$ is even, by Lemma \ref{l:2} we have that $\varepsilon^*=1$. Thus $t^*=\left\lceil\frac{\beta(\varepsilon)}{\lambda}\right\rceil$. 
    \end{itemize}
    In all further cases, $\lambda$ is odd and so the parity of $t$ is important to consider. In what follows, we will write $t^*=\left\lceil\frac{\beta(\varepsilon)}{\lambda}\right\rceil^\textnormal{even}$ or $t^*=\left\lceil\frac{\beta(\varepsilon)}{\lambda}\right\rceil^\textnormal{odd}$, where the notation $a^{\textnormal{even}}$ is the smallest even integer $b$ such that $a\leq b$ (similarly for $a^{\textnormal{odd}}$). For convenience in our current analysis, we will just specify whether $t^*$ is even or odd, but we will summarize the results precisely later.
    \begin{itemize}

    \item Case 3 ($\lambda$ odd, $\tau$ even): By Lemma \ref{l:2} we find that $\varepsilon^*=1$. In this case $L=\tau-2$ is even, so $\beta(\varepsilon)$ is even; we require $t^*$ to be even. 
    \item Case 4 ($\lambda$ odd, $\tau$ odd, $\rho$ even, $\lambda<\tau$): By Lemma \ref{l:2} we find that $\varepsilon^*=1$. In this case $L=\tau-2$ is odd, so $\beta(\varepsilon)$ is odd; we require $t^*$ to be odd. 
    \item Case 5 ($\lambda$ odd, $\tau$ odd, $\rho$ odd, $\lambda<\tau$): We claim that $\varepsilon^*=1.$ Let $(x,y)$ be the minimal point on the line $g(1)$. All points on $g(2)$ can be written as $(x,y)+i(\lambda/2,\lambda/2)+(-\pi/2,\pi/2)$. If the first coordinate is non-negative for some value $i< 0$, then the point $(x,y)-(\lambda,\lambda)$ is also non-negative since $\lambda <\tau\leq \pi$, contradicting minimality of $(x,y)$. Thus $i\ge0$, implying that the second coordinate of all points on $g(2)$ is greater than $y$. Hence $(x,y)$ is the minimal lattice point, and $\varepsilon^*=1$. Since $L=\tau-2$ is odd it follows that $t^*$ must be even. 
    \item Case 6 ($\lambda$ odd, $\tau$ odd, $\rho$ odd, $\tau<\lambda<\pi$): Since $\lambda<\pi$, it follows from the analysis in Case 2c that $\varepsilon^*=1$. Now we have $L=2\tau-2$ which is even, thus $t^*$ must be odd. 
    \item Case 7 ($\lambda$ odd, $\tau$ odd, $\rho=2$, $\tau<\lambda<\pi$): We find that $\pi$ is even, meaning that by Lemma \ref{l:2}, $\varepsilon^*=1$. Since $L=\tau-2$ odd, we find that $t^*$ must be odd. 
    \item Case 8 ($\lambda $ odd, $\tau$ odd, $\rho\neq 2$, $\rho$ even, $\tau <\lambda <\pi$): Since $\pi$ is even, it follows from Lemma \ref{l:2} that $\varepsilon^*=1$. In this case, $L=2\tau-2$ is even, thus $t^*$ must be even. 
    \item Case 9 ($ \lambda $ odd, $\tau$ odd, $\rho=2$, $\lambda >\pi$): As in the previous case, we deduce that $\varepsilon^*=1$. This time $l=\tau-2$ is odd, which means that $t^*$ is odd. 
    \item Case 10 ($ \lambda $ odd, $\tau$ odd, $\rho\neq 2$, $\rho$ even, $\lambda >\pi$): The analysis is identical to Case 8, hence $\varepsilon^*=1$ and $t^*$ must be even. 
    \item Case 11 ($ \lambda $ odd, $\tau$ odd, $\rho$ odd, $\lambda >\pi$): In this case, let us separately analyze the minimal points on $g(1)$ and $g(2)$.
    
    On the line $g(1)$, it follows from parity that $t$ must be odd; the minimal point on this line has a $t$ value of $t_1=\left\lceil\beta(1)/\lambda\right\rceil^\textnormal{odd}$. However, since it cannot be that $\rho \mid \tau$ (else we would have $\sigma=1$), it must be that $\pi \ge 2\tau$. Thus, since $L=2\tau-2>0$, it follows that $0 \leq \beta(1)=\pi-L<\pi$. In particular, we have that $t_1=\left\lceil\beta(1)/\lambda\right\rceil^\textnormal{odd}=\left\lceil\beta(1)/\lambda\right\rceil=1$. 

    It follows by similar reasoning that the $t$ value of the minimal point on $g(2)$ is $t_2=\left\lceil\beta(2)/\lambda\right\rceil^\textnormal{even}$. It is easy to see that $t_1 \leq t_2$, and since they cannot be equal (by parity), we conclude that $t_1<t_2$. Thus, in this case we have that $\varepsilon^*=1$ and $t^*=\left\lceil\beta(1)/\lambda\right\rceil^\textnormal{odd}=1$

\end{itemize}

\begin{theorem}\label{t:valuesP1P2}
    Let $\cP$ be as in Definition \ref{d:lattices}. Then the first lattice point of $\cP$ is given by 
    \[
    P_1=(t^*\lambda+L-\varepsilon^*\pi)/2, ~~~~P_2 = P_1+\varepsilon^*\pi
    \]
    where the values of $t^*,\varepsilon^*$ and $L$ are given in Table \ref{tab:valueP1}.
\end{theorem}

       \begin{table}[h!]
        \centering
\begin{tabular}{||c| c| c |c |c||} 
 \hline
 Case & Conditions &  $L$ & $\varepsilon^*$ &$t^*$\\  
 \hline\hline
 1 & $\lambda$ even, $\rho$ odd 
 & $2\tau-2$ &2
 & $\left\lceil\beta(\varepsilon^*)/\lambda\right\rceil $
\\ 
 \hline
  2 & $\lambda$ even, $\rho$ even
 & $2\tau-2$ & 
 1& $\left\lceil\beta(\varepsilon^*)/\lambda\right\rceil $ \\
 \hline
 3 & \specialcell{$\lambda$ odd, $\tau$ even }
 & $\tau-2$ & 1 &$\left\lceil\beta(\varepsilon^*)/\lambda\right\rceil^\textnormal{even} $
 \\
 \hline

  4 & \specialcell{$\lambda$ odd, $\tau$ odd,\\  $\rho$ even, $\lambda<\tau$ }
 & $\tau-2$  & 1 &  $\left\lceil\beta(\varepsilon^*)/\lambda\right\rceil^\textnormal{odd}$ \\
  \hline
  5 & \specialcell{$\lambda$ odd, $\tau$ odd, \\ $\rho$ odd, $\lambda<\tau$ }
 & $\tau-2$  & 1 & $\left\lceil\beta(\varepsilon^*)/\lambda\right\rceil^\textnormal{even}$ \\
 \hline
 6  & \specialcell{$ \lambda $ odd, $\tau$ odd,\\ $\rho$ odd, $\tau <\lambda <\frac{\tau \rho}{\kappa_2}$}
 & $2\tau-2$ 
 & 1 & $\left\lceil\beta(\varepsilon^*)/\lambda\right\rceil^\textnormal{odd}$ \\
 \hline
 7  & \specialcell{$ \lambda $ odd, $\tau$ odd, \\$\rho=2$, $\tau <\lambda <\frac{\tau \rho}{\kappa_2}$}
 & $\tau-2$ 
 & 1 & $\left\lceil\beta(\varepsilon^*)/\lambda\right\rceil^\textnormal{odd}$ \\
 \hline
  8 & \specialcell{$ \lambda $ odd, $\tau$ odd, $\rho\neq 2$, \\$\rho$ even, $\tau <\lambda <\frac{\tau \rho}{\kappa_2}$}
 & $2\tau-2$ 
 & 1 & $\left\lceil\beta(\varepsilon^*)/\lambda\right\rceil^\textnormal{even} $ \\
 \hline
 9  & \specialcell{$ \lambda $ odd, $\tau$ odd, \\$\rho=2$, $\lambda >\frac{\tau \rho}{\kappa_2}$}
 & $\tau-2$ &1& $\left\lceil\beta(\varepsilon^*)/\lambda\right\rceil^\textnormal{odd}$ \\
 \hline
  10 & \specialcell{$ \lambda $ odd, $\tau$ odd, $\rho\neq 2$,\\ $\rho$ even, $\lambda >\frac{\tau \rho}{\kappa_2}$}
 & $2\tau-2$ &1& $\left\lceil\beta(\varepsilon^*)/\lambda\right\rceil^\textnormal{even} $ \\
 \hline
 11  & \specialcell{$ \lambda $ odd, $\tau$ odd, \\$\rho$ odd, $\lambda >\frac{\tau \rho}{\kappa_2}$}
 & $2\tau-2$ &1&  $\left\lceil\beta(\varepsilon^*)/\lambda\right\rceil^\textnormal{odd}$\\
 \hline
\end{tabular}
        \caption{Values of $L,t^*,\varepsilon^*$ for $(P_1,P_2)$}
        \label{tab:valueP1}
    \end{table}
\begin{remark}
When $\lambda >\pi$ and $\varepsilon^*=1$, we have that $
\lceil\frac{\beta(\varepsilon^*)}{\lambda}\rceil=1
$, and we can write a simpler expression $P_1=(\lambda-\pi+L)/2$.
\end{remark}

\section{Computing the Parameter $c$}\label{s:computec}
In this section we will present explicit formulas to calculate $|\cF_{<k}|$, using the results from the previous sections of the paper. We continue the notations of Sections \ref{s:laticeT} and \ref{s:latticeP}. 
\begin{theorem}\label{t:finalcountingformulas}
    Let $k>1$ be an integer suppose that $\cT_{<k}^1,\cT_{<k}^2,  \cP_{<k}^1, \cP_{<k}^2\neq \emptyset$.  Then each of the following hold: 
\begin{equation}
    \begin{split}
        |\cF_{<k}|& = 2\left( |\cT_{<k}|-|\cP_{<k}|\right ),\\
|\cT_{<k}|&=|\cT_{<k}^1|+|\cT_{<k}^2|,\\
|\cP_{<k}|&=|\cP_{<k}^1|+|\cP_{<k}^2|,\\
    \end{split}
\end{equation}
\begin{equation}
    \begin{split}
               |\cT_{<k}^1|&=\sum_{i=0}^{\left\lceil (k-T_2-\lambda)/\lambda\right\rceil}\left(\min\left\{ \left\lfloor\frac{T_1+i\lambda}{\tau^*}\right\rfloor,  \left\lceil\frac{k-T_2-i\lambda}{\tau^*}-1\right\rceil\right\}+1\right)\\
               |\cT_{<k}^2|&=\sum_{i=0}^{\left\lceil (k-T_2'-\lambda)/\lambda\right\rceil}\left(\min\left\{ \left\lfloor\frac{T_1'+i\lambda}{\tau^*}\right\rfloor,  \left\lceil\frac{k-T_2'-i\lambda}{\tau^*}-1\right\rceil\right\}+1\right)\\
               |\cP_{<k}^1|&=\sum_{i=0}^{\left\lceil (k-P_2-\lambda)/\lambda\right\rceil}\left(\min\left\{ \left\lfloor\frac{P_1+i\lambda}{\pi^*}\right\rfloor,  \left\lceil\frac{k-P_2-i\lambda}{\pi^*}-1\right\rceil\right\}+1\right)\\
               |\cP_{<k}^2|&=\sum_{i=0}^{\left\lceil (k-P_2'-\lambda)/\lambda\right\rceil}\left(\min\left\{ \left\lfloor\frac{P_1'+i\lambda}{\pi^*}\right\rfloor,  \left\lceil\frac{k-P_2'-i\lambda}{\pi^*}-1\right\rceil\right\}+1\right)\\
    \end{split}
\end{equation}
    where $\tau^*=\begin{cases}
    \tau/2 &\textnormal{ if } \tau\textnormal{ is even }\\
\tau &\textnormal{ if } \tau \textnormal{ is odd }
\end{cases}$, similarly for $\pi^*$. 
\end{theorem}

\begin{proof}
    The validity of the first equation is an immediate consequence of Definition \ref{d:lattices} and Equation \ref{eq:sumsublattice}. As described in Remark \ref{r:TandPlattices}, the lattices $\cT$ and $ \cP$ can be written respectively as $\cL_{L,\lambda,\tau}  $ and $\cL_{L,\lambda,\pi} $. The summation formulas then follow from Proposition \ref{p:numberofpoints}. 
\end{proof}
\begin{remark}
    Given the parameters $\lambda,\tau,\rho,\sigma$, one can now compute precisely the value of $\dim(\Hull(\C))$, or equivalently compute the value of $c=\dim C-\dim C\cap C^{\perp_h}=|\cF_{<k}|$, under the assumptions of Lemma \ref{l:countingc}.

    First, refer to Table \ref{tab:valueL} to find the value of $L$. Using Theorems \ref{t:valuesT1T2} and \ref{t:valuesP1P2}, one can then compute the values of $(T_1,T_2), (P_1,P_2)$. The values of $(T_1',T_2'),(P_1',P_2')$ then follow from Lemma \ref{l:firstpointL2}. The formulas in Theorem \ref{t:finalcountingformulas} can then be used to calculate $|\cF_{<k}|$, with reference to Remark \ref{r:whenempty} to check if any lattice is empty. 
\end{remark}

\section{Examples and comparison with current literature}\label{s:comparison}
With respect to the computation of the hull itself, \cite[Table 1]{fangGaloisHullsGRS} shows many of the previous computations for the Galois hulls of MDS codes. The Hermitian hull is a particular case of a Galois hull, but our computations cannot be recovered from the results of \cite[Table 1]{fangGaloisHullsGRS}. For example, we can exactly compute the hull for $1\leq k \leq n $ when $\rho=2$ by Lemma \ref{l:countingc}, and the constructions appearing in  \cite[Table 1]{fangGaloisHullsGRS} have $k<n/2$. In fact, depending on $q$, the maximum $k$ allowed in \cite[Table 1]{fangGaloisHullsGRS} may be much smaller than $n/2$ (in their notation, $q^2=p^h=p^{2e}$ and the upper limit for $k$ is approximately $1+n/p^e$), while we can always compute the Hermitian hull for $1\leq k \leq \lambda \tau$, which might be equal to $n/2$ if we consider $\sigma=2$, for example. 

Regarding EAQMDS codes, by \cite[Thm. 6]{grasslvariableentanglement}, if $q>2$, once we find a code with $\dim \Hull(C)=\ell$, it is always possible to fin a monomially equivalent code $C'$ with $\dim \Hull(C')=\ell'$, for each $\ell'\in \{0,1,\dots,\ell\}$. Thus, when we determine the dimension of the hull with our construction, we also know that there exist GRS codes with lower dimension for their Hermitian hull. In terms of EAQECCs, this implies increasing the parameter $c$. If one starts with a Hermitian self-orthogonal MDS code $C$, then one can derive EAQMDS codes with any $0\leq c \leq \dim C$. For example, this approach is taken in \cite{chenEAQMDS}. However, this limits the minimum distance to, at most, $(n+2)/2$. We do not have this restriction, and thus most of the parameters we obtain cannot be achieved in this way.

In \cite{campionQMDS}, it is shown that the construction we are considering gives new QMDS codes. Similar arguments show that we get new EAQMDS codes. For example, as explained in \cite{campionQMDS}, we may get codes with lengths which are not divisible by $q-1$ and $q+1$, and which do not divide $q^2+1$ nor $q^2-1$. This already discards all the rows of \cite[Table 1]{luoEAQMDSlargedmin} (this is a recent table compiling the known parameters of EAQMDS codes). For example, if we consider $q=29$, $\lambda=28$, $\tau=5$, $\rho=30$ and $\sigma=2$, for $k=28$ we obtain an EAQMDS with parameters $[[280,226,29;2]]_{29}$ (recall Equation (\ref{eq:paramQuant})). The length of this code does not divide $q^2+1$, nor $q^2-1$, and it is not divided by $q-1$ nor $q+1$, which means it is new according to \cite[Table 1]{luoEAQMDSlargedmin}. Another example is given by the parameters $q=11$, $\lambda=5$, $\tau=3$, $\rho=4$ and $\sigma=3$, for $k=9$ we obtain an EAQMDS codes with parameters $[[45,29,10;2]]_{11}$. To finish the comparison, we also consider the recent paper \cite{wanPropagationRulesGRS}. Starting from a QMDS code, \cite[Thm. 9]{wanPropagationRulesGRS} provides a way to obtain EAQMDS with higher minimum distance:

\begin{theorem}
For $q>2$, assume there is an $[[n,n-k-l,k+1;k-l]]_q$ EAQMDS code constructed with the Hermitian construction \ref{t:hermitian} from an $[n,k]_{q^2}$ GRS code with $l$-dimensional Hermitian hull, where $0\leq 2k\leq n $ and $0\leq l \leq k$. Then for any integer $0\leq i \leq \min \{l,q^2+1-n,n-2k\}$ and $0\leq s \leq l-i$, there is an $[[n,n-k-i-s,k+i+1;k+i-s]]_q$ EAQMDS code. 
\end{theorem}

Because of the limitation on the parameters, the resulting code will always have $c=k+i-s\geq k+2i-l$. If we want to increase the minimum distance by $i$, then the parameter $c$ will also increase by, at least, $2i$. However, we have many instances in which we can increase the minimum distance without increasing the parameter $c$ (see Table \ref{T:newparameters}). Thus, one cannot use this result to derive our parameters from those in \cite{campionQMDS}. For example, in \cite{campionQMDS} the authors obtain a code with parameters $[[45,33,7;0]]_{11}$. Using the previous result, one can obtain the codes $[[45,39-i-s,7+i;6+i-s]]_{11}$, $0\leq i \leq 6$, $0\leq s \leq 6-i$. With minimum distance $10$, we have $i=3$, and we obtain the codes $[[45,36-s,10;9-s]]_{11}$, $0\leq s \leq 3$. The one with lowest entanglement has parameters $[[45,33,10;6]]_{11}$, which can be derived from our $[[45,29,10;2]]_{11}$ using the usual propagation rules (as stated before, we can decrease the dimension of the hull, which implies increasing $c$ and, consequently, the dimension of the quantum code). By a similar reasoning, one cannot obtain the parameters of our codes simply by using some of the propagation rules from \cite{wanPropagationRulesGRS} (or those from \cite{grasslvariableentanglement}). 

{\small
\begin{table}[ht]
\caption{Parameters of some new EAQMDS codes.}\label{T:newparameters}
\centering
\begin{tabular}{c|c}
 \hline 
 $k$&Parameters \\
  \hline \hline
  &$q=11, \lambda=5, \tau=3, \rho=4, \sigma=3$\\ 
  \hline
$8$& $[[45, 31, 9; 2]]_{11}$  \\

$9$& $[[45, 29,10; 2]]_{11}$  \\
$10$& $[[45, 29, 11; 4]]_{11}$ \\
$11$& $[[45, 29, 12; 6]]_{11}$  \\
$12$& $[[45, 29, 13; 8]]_{11}$   \\
$13$& $[[45, 27, 14; 8]]_{11}$   \\
$14$& $[[45, 25, 15; 8]]_{11}$   \\
$15$& $[[45, 25, 16; 10]]_{11}$   \\

  \hline \hline
  &$q=83,\lambda=41,\tau=6,\rho=84,\sigma=2$\\ 
  \hline
$ 48 $& $[[ 492 , 398 , 49 ; 2 ]]_{ 83 }$\\
$ 49 $& $[[ 492 , 396 , 50 ; 2 ]]_{ 83 }$\\
$ 50 $& $[[ 492 , 396 , 51 ; 4 ]]_{ 83 }$\\
$ 51 $& $[[ 492 , 394 , 52 ; 4 ]]_{ 83 }$\\
$ 52 $& $[[ 492 , 392 , 53 ; 4 ]]_{ 83 }$\\
$ 53 $& $[[ 492 , 392 , 54 ; 6 ]]_{ 83 }$\\
$ 54 $& $[[ 492 , 390 , 55 ; 6 ]]_{ 83 }$\\
$ 55 $& $[[ 492 , 388 , 56 ; 6 ]]_{ 83 }$\\
$ 56 $& $[[ 492 , 388 , 57 ; 8 ]]_{ 83 }$\\
$ 57 $& $[[ 492 , 386 , 58 ; 8 ]]_{ 83 }$\\
$ 58 $& $[[ 492 , 384 , 59 ; 8 ]]_{ 83 }$\\
$ \vdots $& $\vdots$\\
$ 233 $& $[[ 492 , 228 , 234 ; 202 ]]_{ 83 }$\\
$ 234 $& $[[ 492 , 228 , 235 ; 204 ]]_{ 83 }$\\
$ 235 $& $[[ 492 , 228 , 236 ; 206 ]]_{ 83 }$\\
$ 236 $& $[[ 492 , 228 , 237 ; 208 ]]_{ 83 }$\\
$ 237 $& $[[ 492 , 228 , 238 ; 210 ]]_{ 83 }$\\
$ 238 $& $[[ 492 , 228 , 239 ; 212 ]]_{ 83 }$\\
$ 239 $& $[[ 492 , 228 , 240 ; 214 ]]_{ 83 }$\\
$ 240 $& $[[ 492 , 228 , 241 ; 216 ]]_{ 83 }$\\
$ 241 $& $[[ 492 , 228 , 242 ; 218 ]]_{ 83 }$\\
$ 242 $& $[[ 492 , 228 , 243 ; 220 ]]_{ 83 }$\\
$ 243 $& $[[ 492 , 228 , 244 ; 222 ]]_{ 83 }$\\
$ 244 $& $[[ 492 , 228 , 245 ; 224 ]]_{ 83 }$\\
$ 245 $& $[[ 492 , 228 , 246 ; 226 ]]_{ 83 }$\\
$ 246 $& $[[ 492 , 228 , 247 ; 228 ]]_{ 83 }$\\
\end{tabular}
\begin{tabular}{c|c}
 \hline 
 $k$&Parameters \\
  \hline \hline
&$q=29,\lambda=28,\tau=5,\rho=30,\sigma=2$\\ 
  \hline
$ 25 $& $[[ 280 , 232 , 26 ; 2 ]]_{ 29 }$\\
$ 26 $& $[[ 280 , 230 , 27 ; 2 ]]_{ 29 }$\\
$ 27 $& $[[ 280 , 228 , 28 ; 2 ]]_{ 29 }$\\
$ 28 $& $[[ 280 , 226 , 29 ; 2 ]]_{ 29 }$\\
$ 29 $& $[[ 280 , 226 , 30 ; 4 ]]_{ 29 }$\\
$ 30 $& $[[ 280 , 224 , 31 ; 4 ]]_{ 29 }$\\
$ 31 $& $[[ 280 , 222 , 32 ; 4 ]]_{ 29 }$\\
$ 32 $& $[[ 280 , 220 , 33 ; 4 ]]_{ 29 }$\\
$ 33 $& $[[ 280 , 218 , 34 ; 4 ]]_{ 29 }$\\
$ 34 $& $[[ 280 , 216 , 35 ; 4 ]]_{ 29 }$\\
$ 35 $& $[[ 280 , 214 , 36 ; 4 ]]_{ 29 }$\\
$ 36 $& $[[ 280 , 212 , 37 ; 4 ]]_{ 29 }$\\
$ 37 $& $[[ 280 , 210 , 38 ; 4 ]]_{ 29 }$\\
$ 38 $& $[[ 280 , 210 , 39 ; 6 ]]_{ 29 }$\\
$ 39 $& $[[ 280 , 208 , 40 ; 6 ]]_{ 29 }$\\
$ 40 $& $[[ 280 , 206 , 41 ; 6 ]]_{ 29 }$\\
$ 41 $& $[[ 280 , 204 , 42 ; 6 ]]_{ 29 }$\\
$ 42 $& $[[ 280 , 202 , 43 ; 6 ]]_{ 29 }$\\
$ 43 $& $[[ 280 , 202 , 44 ; 8 ]]_{ 29 }$\\
$ 44 $& $[[ 280 , 200 , 45 ; 8 ]]_{ 29 }$\\
$ \vdots $& $\vdots$\\
$ 127 $& $[[ 280 , 100 , 128 ; 74 ]]_{ 29 }$\\
$ 128 $& $[[ 280 , 100 , 129 ; 76 ]]_{ 29 }$\\
$ 129 $& $[[ 280 , 100 , 130 ; 78 ]]_{ 29 }$\\
$ 130 $& $[[ 280 , 100 , 131 ; 80 ]]_{ 29 }$\\
$ 131 $& $[[ 280 , 98 , 132 ; 80 ]]_{ 29 }$\\
$ 132 $& $[[ 280 , 96 , 133 ; 80 ]]_{ 29 }$\\
$ 133 $& $[[ 280 , 94 , 134 ; 80 ]]_{ 29 }$\\
$ 134 $& $[[ 280 , 92 , 135 ; 80 ]]_{ 29 }$\\
$ 135 $& $[[ 280 , 90 , 136 ; 80 ]]_{ 29 }$\\
$ 136 $& $[[ 280 , 90 , 137 ; 82 ]]_{ 29 }$\\
$ 137 $& $[[ 280 , 90 , 138 ; 84 ]]_{ 29 }$\\
$ 138 $& $[[ 280 , 90 , 139 ; 86 ]]_{ 29 }$\\
$ 139 $& $[[ 280 , 90 , 140 ; 88 ]]_{ 29 }$\\
$ 140 $& $[[ 280 , 90 , 141 ; 90 ]]_{ 29 }$\\
\end{tabular}
\end{table}
}



\bibliographystyle{abbrv}

\end{document}